\documentclass[a4paper,12pt,leqno]{article}

\usepackage{amssymb,amsmath} 
\usepackage{amsthm}
\usepackage[author-year]{amsrefs}
\usepackage{xy}
\xyoption{all}
\usepackage{graphicx}
\usepackage{hyperref}
\usepackage[text={15.5cm,25cm}]{geometry}

\newtheorem*{xdefn}{Definition}
\newtheorem*{xexm}{Example}
\newtheorem*{lemma*}{Lemma}
\newtheorem{lemma}{Lemma}[section]
\newtheorem*{theorem*}{Theorem}
\newtheorem*{corollary*}{Corollary}
\newtheorem{theorem}[lemma]{Theorem}
\newtheorem{proposition}[lemma]{Proposition}
\newtheorem*{conjecture*}{Conjecture}
\newtheorem*{proposition*}{Proposition}
\newtheorem*{xrem}{Remark}
\newtheorem*{thesis*}{Thesis}

\newtheorem*{principle*}{Axiom}

\newenvironment{definition*}{\begin{xdefn}\em}{\end{xdefn}}
\newenvironment{example*}{\begin{xexm}\em}{\end{xexm}}
\newenvironment{remark*}{\begin{xrem}\em}{\end{xrem}}

\newcommand\V{\bigvee}
\newcommand\ie{i.e.}
\newcommand\eg{e.g.}

\newcommand\cf{\textrm{cf.}}

\newcommand\topology{\operatorname{\Omega}}

\newcommand\Q{\mathcal Q}

\newcommand\B{\mathcal B}
\newcommand\redquale{\mathbf{red}}

\newcommand\bluequale{\mathbf{blue}}
\newcommand\impquale{\boldsymbol{0}}
\newcommand\C{\mathcal R}

\begin{document}

\title{A physical approach to qualia and the emergence of conscious observers in qualia space\thanks{Work funded by FCT/Portugal through project UID/04459/2025.}}

\author{Pedro Resende\\
~\\
\begin{minipage}{\textwidth}
\it\normalsize
\center Center for Mathematical Analysis, Geometry and Dynamical Systems,\\
Department of Mathematics, Instituto Superior T\'{e}cnico,
University of Lisbon,\\
Av.\ Rovisco Pais 1, 1049-001 Lisboa, Portugal\\
~\\
E-mail: {\sf pedro.m.a.resende@tecnico.ulisboa.pt}
\end{minipage}
}

\date{~}

\maketitle

\vspace*{-1cm}
\begin{abstract}
I propose that qualia are physical because they are directly observable, and revisit the contentious link between consciousness and quantum measurements from a new perspective---one that does not rely on observers or wave function collapse but instead treats physical measurements as fundamental in a sense resonant with Wheeler's it-from-bit. Building on a mathematical definition of measurement space in physics, I reinterpret it as a model of qualia, effectively equating the measurement problem of quantum mechanics with the hard problem of consciousness. The resulting framework falls within panpsychism, and offers potential solutions to the combination problem. Moreover, some of the mathematical structure of measurement spaces, taken for granted in physics, needs justification for qualia, suggesting that the apparent solidity of physical reality is deeply rooted in how humans process information.
\end{abstract}

{\small \tableofcontents}

\section{Introduction}

The \emph{hard problem of consciousness}~\cites{Chalmers1,Chalmers-book2} is inseparable from the word ``qualia,'' originally coined by Lewis~\ycite{MindWorldOrder}. A quale is the quality of an indivisible moment of experience. Given the prevailing consensus that consciousness is extended in time~\cite{KentWittmann01}, qualia should persist in time or indeed be regarded as \emph{generators} of subjective time: the subjective quality attached to a moment of experience carries an indissociable ``feeling'' of elapsed time. As examples of qualia, consider an animal's raw experience of feeling hot, cold, or hungry; but also sophisticated AHA! moments like a mathematician's sudden discovery of the missing link in a mathematical proof, or Descartes realizing that he thinks, therefore he exists.

The role of qualia in science remains unclear. Regarding physics as the theory of observable phenomena in the world, consciousness must belong to physics because, by definition, \emph{qualia are directly observable}---whether by naive introspection, or by skilled methods of meditators. 
Hence, the quest for understanding qualia has multifaceted importance. Besides philosophical relevance, it has practical consequences for our ability to control consciousness-related phenomena in medical practice, or to assess whether a robot is conscious, etc.

However, the scientific study of qualia is challenging because science has evolved seeking objective conclusions that circumvent subjective experience of experimenters. Regardless, the science of consciousness has been maturing in the past decades. Bridges between neuroscience and subjective experience are given by first-person reports of people undergoing fMRI scans---see, \eg,~\ocite{GWT11} and references therein---and such experiments satisfy one of the fundamental tenets of science: they can be replicated. But they are limited to the consciousness of humans with verbal capabilities.

Going beyond is harder. Scientific confidence regarding consciousness is mostly limited to animals, as symbolically proclaimed by The Cambridge Declaration on Consciousness~\cite{CambridgeDeclarationConsciousness} and recently by The New York Declaration on Animal Consciousness~\cite{NYDeclarationConsciousness}. This excludes artificial systems, no matter how ``human'' they look, and even complex biological systems like plants, trees, or forests, despite their evidence of complex behavior and communication capabilities.

Any scientific endeavor must be phrased clearly enough to enable experimental testing. In this spirit, several models of consciousness inspired by human neurobiology have been developed and their predictions can be compared~\cite{SethBayne2022}. Two of the best known are \emph{integrated information theory} (IIT)~\cites{Ton04,IIT3,IIT4} and \emph{global workspace theory} (GWT)~\cites{Baars, GWT11, GNWT20}. The former is a mathematical theory focused on quantifying phenomenological experience. The latter is more function-oriented and has inspired a mathematical model called the \emph{Conscious Turing Machine (CTM)}~\cite{CTM}, which has interesting features like the ability to reproduce effects of certain neurological disorders.

In science the need to be rigorous is a road that leads to mathematics. For example, IIT and the CTM are mathematical. Other mathematical models have been proposed, \eg, mathematical spaces that describe structural aspects of phenomenology---see~\ocite{KleinerPhD} and references therein. They are motivated by the pursuit of conceptual clarity, a necessary condition for enabling experimental testing, even if for some models no immediate experimental procedures are in sight. This type of research has been expanding for several years; it is now called \emph{mathematical consciousness science}, and has motivated the creation of the Association for Mathematical Consciousness Science (AMCS).\footnote{https://amcs-community.org/}

But the hard problem of consciousness, particularly its connection (or lack thereof) to physics, remains unresolved. This issue sparks intense debate~\cite{ConscQM} and often brings us to the \emph{measurement problem of quantum mechanics}, which became notorious after von Neumann~\ycite{vonNeumann1955}*{Ch.\ VI}  proposed that measurements depend on subjective perception, despite such a notion being undefined in physics. This created the daunting question of which systems qualify as ``observers,'' and wrecked conceptual havoc in a fundamental pillar of modern physics. As John Bell~\ycite{Bell90} humorously wrote,
\begin{quote}
``What exactly qualifies
some physical systems to play the role of `measurer'? Was the wavefunction
of the world waiting to jump for thousands of millions of years until a single-celled living creature appeared? Or did it have to wait a little longer, for some
better qualified system...\ with a PhD?''
\end{quote}
The difficulties persist, along with a steady proliferation of interpretations and modifications of quantum mechanics~\cites{klaas,AdamBecker,Mer12}.

A class of strong contenders consists of \emph{objective collapse models} that modify the Schr\"odinger equation by adding stochastic terms that control the rate of collapse of wave functions. The collapse term of one of the earliest and best known models~\cite{GRW86} depends on two new physical constants, but other proposals attempt to relate collapse to consciousness, \eg, the approach of Hameroff and Penrose~\ycite{HamPen14}, whose collapse rate is determined by gravity, and approaches whose rate is controlled by quantum versions of IIT~\cites{KreRan15,OkoSeb18,ChalmersMcQueen}. These proposals are modifications of quantum mechanics and need to be experimentally tested.

We can rephrase the measurement problem as the question of how classical information arises from a quantum system. For instance, if a qubit is measured along a specified direction the result is a bit of classical information, say 0 or 1. The measurement is a moment of \emph{creation} of classical information with no analogue in standard deterministic digital computers, whose sole ability is to transform or delete bits. This view led Wheeler~\ycite{itfrombit} to propose that creation of classical information is the manifestation of reality, which he reinforced by coining the famous expression ``it-from-bit.'' We are thus led to viewing creation of classical information as a fundamental ingredient of conscious experience---concurring with the strong free will theorem of~\ocite{ConKoc09}, which suggests that measurements are fundamental choices made by objects being measured, indistinguishably from how a human chooses the details of an experimental apparatus. Importantly, if-from-bit does \emph{not} mean that consciousness emerges from ``information processing'' as in a typical digital computer. It also does not support solipsist claims that consciousness subsumes matter. For example, measuring the spin of an electron along an axis creates a bit of information about the electron, but it does \emph{not} create the electron.

This paper explores an approach to qualia based on mathematical spaces developed to describe quantum measurements~\cites{fop,msnew}. This framework equates measurements with the classical information they produce, offering a mathematical formulation of it-from-bit. It suggests that qualia and measurements are two sides of the same coin, implying that the hard problem of consciousness is inseparable from the measurement problem. While this perspective may indicate that physical measurements and qualia are fundamental in nature, at least it abstracts away from specifics like wave function collapse, allowing new mathematical structure to emerge---which, besides improving conceptual clarity, helps formulate questions for experimental psychology.

This approach to qualia is panpsychist, addressing the subject combination problem~\cite{Chalmers-panpsychism} through how ``observers'' arise within spaces of measurements. Moreover, observers are naturally equipped with a geometric vision of reality, provided they have: (i)~enough cognitive capabilities for performing appropriate computations; and (ii)~no direct awareness of quantum interference.

Different proposals linking qualia to quantum mechanics exist, for instance associating qualia to quantum states~\cite{DArianoFaggin}. Despite having similar names, our spaces are very different from the qualia space of Stanley~\ycite{Stanley}, both in terms of rationale and mathematical structure.

Although this paper is a contribution to mathematical consciousness science, its mathematical prerequisites are minimal---basic  notation from naive set theory suffices, \eg, membership $x\in X$, set containment $X\subset Y$, intersections $X\cap Y$ and unions $X\cup Y$, and their infinitary forms $\bigcap_{i\in I} U_i$ and $\bigcup_{i\in I} U_i$. Additional mathematical concepts will be introduced as needed.

\section{Rationale}\label{sec:rationale}

Any physical measurement, whatever collapse is driven by, causes classical information to be stored in measuring devices, photographic plates, brain synapses, etc. Even under the many worlds interpretation~\cite{Everett}, where only decoherence produces an apparent collapse~\cite{decoherencebook}, classical information arises from tracing out the environment. Thus, we define \emph{abstract measurements} as equivalence classes of physical experiments that yield the same classical information. For example, the two-slit experiment and Wheeler's delayed-choice variant---see, \eg, \ocite{GreensteinZajonc}*{p.\ 39--44}---are indistinguishable by the classical information they yield: knowledge of a particle's path erases interference, while its absence preserves it.

These ideas led to the notion of a \emph{measurement space}~\cites{fop,msnew}, a mathematical space whose points represent abstract measurements and whose open sets (\cf\ section~\ref{sec:qualiaconcepts}) correspond to physically observable properties. Measurement spaces also contain structures that represent time, causality, and repeatability of measurements.

This paper applies the measurement space framework to qualia, abstracting away from brain mechanisms and instead characterizing qualia by the classical information they generate. Qualia are modeled as points in a space whose open sets represent physically stored finite amounts of classical information, such as long-term memories in a brain. These finite information units constitute communicable knowledge and correspond to the definition of a \emph{pure concept}~\cite{MindWorldOrder}, herein just called a \emph{concept}.

We explore the analogy between measurements and qualia, reinterpreting measurement spaces as qualia spaces. Importantly, measurement spaces do not need pre-existing observers; instead, ``classical observers'' emerge \emph{within} these spaces. Similarly, qualia do not require individuated conscious entities. While we speak of Alice's or Bob's qualia, the experience of being Alice or Bob emerges from complex and coherent subspaces of qualia associated with dynamic physical structures. This perspective aligns with neuroscience, which views the ``self'' as a construct of the brain, and it offers insights into the subject combination problem.

Measurement and qualia spaces model the creation of classical information. Whether they abstract a more fundamental theory, \eg, involving objective collapse, or just convey that information is fundamental, remains to be seen. Either way, this approach provides a mathematical framework free from the conceptual difficulties of quantum measurements, potentially addressing both the hard problem of consciousness and the measurement problem: measurements are physical processes through which qualia arise, forming the seeds of experience.

A key issue is how much of the \emph{mathematical structure} of qualia spaces is fundamental. For example, measurement spaces have an operation called composition, which reflects time and causality: performing $\beta$ and then $\alpha$ is a new measurement $\alpha\beta$. However, when applying this to qualia, we must explain how the experience of $\alpha\beta$ arises as an indivisible unit. This requires sufficient cognitive structure to support the detection of a composition and the subjective experience it entails---\cf~section~\ref{sec:cognitivecaps}.

Testing whether measurement structures can be justified in terms of cognitive abilities requires psychophysical experiments. This paper serves as an initial step toward such research. Notably, it is unclear whether such experiments are supposed to detect an objective reality or merely reveal how brains process information.

These ideas resonate with the conscious realism program of~Hoffman, Prakash, et al---see~\ocite{DonaldHoffman}. While our spaces share similarities with part of their framework, an important motivation behind measurement spaces is to reconstruct quantum mechanics, for which additional structure related to dynamics and quantum probability, currently being studied, needs to be added.

The remainder of this paper explores qualia by examining some mathematical structure and properties of measurement spaces, focusing on concepts (a \emph{topology} in the mathematical sense) and the operations of \emph{composition} (related to time) and \emph{disjunction} (related to logical abstraction). We also examine how ``classical observers'' emerge within the space of qualia, offering an approach to the subject combination problem in panpsychism. The presented mathematical structures suggest questions that can be tested on humans by means of appropriate psychophysical experiments. Hints of these will be given.

\section{Qualia and concepts}\label{sec:qualiaconcepts}

\paragraph{Qualia.}

Let us begin with Nagel's~\ycite{Nagel} question: ``What is it like to be a bat?'' Consider variations of this for Bob: ``What is it like to be Bob seeing a flash of red light?''; or ``Bob sunbathing on the beach?''; ``Bob in a bad mood?''; ``Bob startled by thunder?''; or something sophisticated like ``What is it like to be Bob suddenly thinking that he is an independent being thinking that he is an independent being?''; or ``Bob thinking that he thinks, therefore he exists?''. Each answer is a quale, \emph{possessing the indivisibility characteristic of conscious experience}.

It will be seen that a property of qualia spaces is that qualia are partially ordered by \emph{specificity}. For example, ``Bob sees red light'' is more specific than ``Bob sees light.'' This gives rise to diagrams like that of Fig.~\ref{fig:orderBob}, where specificity increases as we move down, and $\impquale$ is the most specific quale of all, a mathematical convention called the \emph{impossible quale}.

\begin{figure}[h!]
\[
\xymatrix{
&\text{Being Bob}\ar@{-}[dl]\ar@{-}[dr]\\
\text{Bob sees light}\ar@{-}[d]&&\text{Bob has a thought}\ar@{-}[d]\\
\text{Bob sees red light}\ar@{-}[dr]&&\text{Bob thinks about pizza}\ar@{-}[dl]\\
&\impquale
}
\]
\caption{A subset of Bob's qualia partially ordered by specificity.}\label{fig:orderBob}
\end{figure}

The existence of the least specific quale ``Being Bob'' can be interpreted as Bob's sense of self, his most abstract experience---more abstract than all the ``sophisticated qualia'' listed above.
An organism capable of experiencing such ``higher qualia'' needs powerful cognitive capabilities and is likely perceived as ``more conscious'' than another whose qualia are direct sensory experiences, such as perhaps an insect, in Fig.~\ref{fig:orderInsect} depicted without a sense of self:

\begin{figure}[h!]
\[
\xymatrix{
\text{Insect sees light}\ar@{-}[d]&&\text{Insect is restless}\ar@{-}[d]\\
\text{Insect sees red light}\ar@{-}[dr]&&\text{Insect is hungry}\ar@{-}[dl]\\
&\impquale
}
\]
\caption{Some qualia of an imaginary insect.}\label{fig:orderInsect}
\end{figure}

Contrasting with IIT, the ``amount'' of consciousness is conveyed by how high an organism's qualia sit in the specificity order, rather than being assigned a numerical value---not all qualia can be compared, since the order is partial. But, regardless of how high it sits, each quale is an indivisible conscious experience---lying above $\alpha$ and $\beta$ does not mean that a quale has ``parts'' $\alpha$ and $\beta$. And, while Bob can have more sophisticated experiences than those of an insect, the \emph{nature} of their consciousness is the same: it is made of qualia.

These ideas will be revisited in section~\ref{sec:agents} when describing complex agents in qualia space and the subject combination problem of panpsychism.

\paragraph{The space of qualia.}

A measurement space $M$ contains points $\alpha\in M$, representing abstract physical measurements, plus certain subsets $U\subset M$, corresponding to physical properties that can be recorded using finite time and energy. The meaning of the assertion $\alpha\in U$ is that $U$ can be the property recorded when the measurement $\alpha$ is performed. For instance, let $\alpha$ be a measurement of a spin-1/2 particle using a Stern--Gerlach apparatus~\cite{GreensteinZajonc} with which we observe an upwards or a downwards deflection of the particle. Denoting the corresponding measurable properties as $U^\uparrow$ and $U^\downarrow$, we have $\alpha\in U^\uparrow\cap U^\downarrow$, since both can be recorded---albeit not simultaneously.

The collection of such subsets satisfies:
\begin{enumerate}
\item $M\in\topology(M)$.
\item $U,V\in\topology(M)$ implies $U\cap V\in\topology(M)$.
\item $\bigcup_{i\in I} U_i\in\topology(M)$ for any indexed family $(U_i)_{i\in I}$\quad (hence $\emptyset\in \topology(M)$).
\end{enumerate}
Mathematically, such a collection is called a \emph{topology} on $M$, denoted $\topology(M)$, and the elements $U\in\topology(M)$ are the \emph{open sets} of $M$. The set $M$ together with its open sets is a \emph{topological space}~\cite{Munkres}.

Logical \emph{conjunctions} of properties correspond to intersections $U\cap V$ (as in the above spin example). In the absence of other assumptions, the canonical way to test the assertion $\alpha\in U\cap V$ is to run the measurement $\alpha$ twice, recording $U$ and $V$ in succession. However, an infinite intersection $\bigcap_i U_i$ cannot be explained like this. In contrast, arbitrary disjunctions, represented by unions $\bigcup_i U_i$, are testable by verifying a single condition $\alpha\in U_i$. This asymmetry explains why measurable properties form a topology: open sets are a mathematical expression of the finite means available to perform measurements---an idea borrowed from computer science~\cite{topologyvialogic}.

Translating to qualia and concepts, let $\Q$ be the set whose points are all the possible qualia, independently of specific observers or locations, equipped with a topology $\topology(\Q)$ whose open sets correspond to concepts. If $\alpha\in U$, we say that $U$ can be the concept recorded when experiencing $\alpha$; or, equivalently, $\alpha$ can \emph{trigger} $U$.

For instance, let $U$ be Alice's concept of the color red. As a subset of $\Q$, $U$ contains all the qualia associated with Alice's red experiences: seeing a sunset, a red laser beam, etc. These qualia can induce the recording of $U$ in Alice's synapses.

For conjunctions of concepts, consider the following thought experiment: Alice is exposed twice to a stimulus that she verbally acknowledges as a repetition of the same subjective experience $\alpha$; but, each time, a lab technician observes distinct activations via brain imaging, corresponding to concepts $U$ and $V$, thus establishing $\alpha\in U\cap V$.

The set $\Q$, equipped with $\topology(\Q)$, is the \emph{space of qualia}. The interpretation of concepts as open sets can also be interpreted as follows: when a quale $\alpha$ triggers Alice's concept of red ($\alpha\in U$), she is certain of experiencing redness. But if $\alpha\notin U$, she may be unsure of having experienced redness because, in topological terms, each $\alpha\in U$ is an \emph{interior point} of $U$, whereas $\alpha\notin U$ may be an \emph{exterior} or a \emph{boundary} point. Hence, $U$ is like a \emph{semi-decidable property} in computer science~\cite{Smyth83}.

\paragraph{Interdependence of qualia and concepts.}

At first glance, the measurement space model of qualia seems to imply dualism, as it distinguishes qualia (mental phenomena) from concepts (physical entities). However, this is not true, as I will explain.

We require $\Q$ to have the following property inherited from measurement spaces: any qualia that trigger exactly the same concepts must be identical. Mathematically, this means that $\Q$ is a space of \emph{type $T_0$}, or a \emph{$T_0$-space}. Denoting by $N(\alpha)$ the set of concepts that $\alpha$ can trigger, the $T_0$ property means that $N(\alpha)=N(\beta)$ implies $\alpha=\beta$ for all $\alpha,\beta\in\Q$. The notation $N(\alpha)$ is motivated by common topological terminology, namely each $U\in N(\alpha)$ is called a \emph{neighborhood} of $\alpha$.

An evident causal relationship is that qualia influence concepts. For instance, human subjective experiences shape brain synapses that encode concepts. However, the reverse also holds. Consider newborn Alice experiencing light for the first time. Suppose, for the sake of the argument, that her initial perception was pure brightness, without color. Over time, after exposure to flashes of different wavelengths, her synapses adapted, allowing her to distinguish colors.
The modified synaptic structures encode her concepts of red, blue, etc., even though her initial visual stimuli lacked color qualia. This suggests that Alice's ability to experience color emerged as her brain developed the corresponding concepts.

Thus, qualia and concepts are interdependent: qualia arise only when the relevant concepts exist, making the mental and physical worlds inseparable.

\paragraph{Logical abstraction of qualia.}

\begin{figure}[h!]
\begin{center}\includegraphics[width=0.5\textwidth]{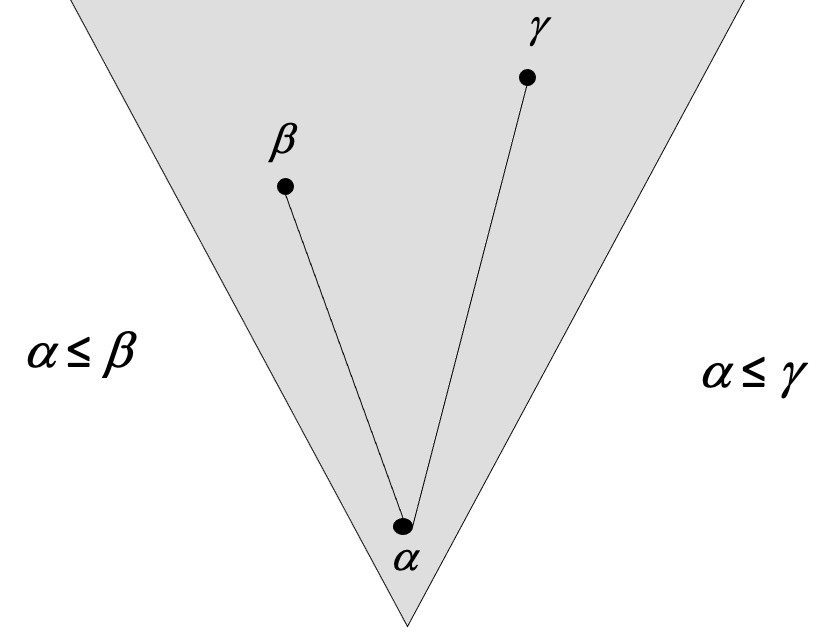}
\caption{The concept $U$, represented by a triangle, is upwards closed in the specialization order of $\Q$.}\label{fig:order}\end{center}
\end{figure}

Some qualia are more specific than others. For example, seeing a flash of light is less specific than seeing a flash of red light, which in turn is less specific than seeing a flash of intense red light. Given two qualia, $\alpha$ and $\beta$, we say that $\alpha$ is \emph{more specific} than $\beta$, and write
\[
\alpha\le \beta,
\]
if every concept $U$ that $\alpha$ can trigger can also be triggered by $\beta$; that is, $N(\alpha)\subset N(\beta)$. This relation, known as the \emph{specialization order} of the  space $\Q$, reflects the idea that $\alpha$ is more constrained, being associated with fewer potential concepts, whereas $\beta$ is less determined. Consequently, any concept $U$ is upwards closed in this order: if $\alpha\in U$ and $\alpha\le\beta$, then $\beta\in U$ (Fig.~\ref{fig:order}).

Importantly, an inequality $\alpha\le\beta$ does not entail conscious awareness that $\alpha$ is more specific than $\beta$, \ie, that ``$\alpha$ implies $\beta$.'' Despite the section title, the $\le$ relation does not assume an underlying cognitive ability for logic and deduction. Rather, it is a feature of any topological space. However, below we will see that this relation is a basis for abstract descriptions of cognitive abilities related to logic.

\section{Cognitive abilities}\label{sec:cognitivecaps}

\paragraph{Subjective time.}

Any measurement space has an operation of \emph{composition} that represents a primitive notion of time: 
$\alpha\beta$ signifies ``$\beta$ and then $\alpha$'' or, more restrictively, ``$\beta$ and only then $\alpha$,'' implying a causal dependence. In quantum mechanics, measuring an electron's spin along the $x$-axis, followed by a measurement along the $z$-axis, is an example of a causal dependence.

Applying this idea to qualia, if 
$\alpha$
and 
$\beta$
are experiences, what does the new quale
$\alpha\beta$ represent? A plausible interpretation, given sufficient cognitive ability, including memory, is that 
$\alpha\beta$ is the AHA! realization that experiences 
$\beta$
and 
$\alpha$ occurred sequentially.

Regardless of how objective time is defined, the duration between experiences is irrelevant. The composition $\alpha\beta$ means that $\beta$ and $\alpha$ occurred consecutively, with no intervening experiences. For example, a patient undergoing surgery may recall the last moment before anesthesia as if it immediately preceded the moment of waking up, despite hours passing in between. Since no \emph{subjective time} existed while unconscious, qualia composition is \emph{associative}: for any qualia 
$\alpha$, $\beta$, $\gamma$, we make no distinction between
$(\alpha\beta)\gamma$ and $\alpha(\beta\gamma)$---both represent the experience $\gamma$, followed by $\beta$, then $\alpha$.

This perspective suggests that subjective time can be regarded as a \emph{measure of how many qualia occur}, mirroring ideas in psychophysical research as illustrated by the following passage of~\ocite{subjectivetime}:

\begin{quote}
``Imagine that one bit of internal information processed is interpreted as one unit of objective time having passed. When the rate of internal information processing suddenly goes up to two bits per unit of objective time (as when one pays more attention because of an imminent crash into another car), a counter would count more bits. If the assessment of duration by the brain is the result of the output of such a counter, it would come to the wrong conclusion that more objective time had passed, creating the illusion that time and motion had slowed
down.''
\end{quote}

These ideas reinforce the view that qualia are \emph{extended in time}, meaning each quale is indissociable from an underlying experience of duration. Consequently, a composition of  qualia $\alpha_1\cdots\alpha_n$ should produce a perceived duration corresponding to the sum of the subjective durations of the qualia $\alpha_i$. This is consistent with phenomenological and neuroscientific approaches to consciousness~\cite{KentWittmann01}.

A question is whether the AHA! moments associated with compositions $\alpha\beta$ can be detected in human brains and how they relate to current knowledge about neural correlates of subjective time. Perhaps this can be tested via brain imaging conjugated with first-person reports, using masking to distinguish conscious from unconscious detections of compositions. We anticipate such experimental research will be conducted elsewhere.

Additional structure related to composition exists in measurement spaces but will not be examined here.

\paragraph{Logical disjunctions of qualia.}

Returning to the discussion in section~\ref{sec:qualiaconcepts}, let us explore how the specialization order of $\Q$ provides a foundation for rudimentary logical reasoning.

Given two qualia $\alpha$ and $\beta$, we define their \emph{disjunction} $\alpha\vee \beta$ to be the most specific quale which is less specific than both $\alpha$ and $\beta$ (should it exist). In other words, $\alpha\vee \beta$ is the \emph{supremum} of $\alpha$ and $\beta$ in the specialization order.

A natural interpretation of $\alpha\vee\beta$ emerges from the role of repetition in learning, assuming sufficient cognitive capacity, perhaps through neural mechanisms like those of~\ocite{Buchsbaumetal2015}.
For example, consider qualia $\redquale$ and $\bluequale$, representing the experience of watching a flash of red light and a flash of blue light, respectively. Suppose a flashlight emits either red or blue light at random, each time we press a button. After several presses of the button, the realization that ``either red or blue'' will appear is an AHA! moment---a more abstract quale that we denote as $\redquale\vee\bluequale$.

This is the least amount of information derivable from repetition. More advanced cognitive abilities---such as counting occurrences---could lead to probabilistic reasoning rather than only logical disjunction. This distinction is significant, as logical reasoning and basic arithmetic in humans rely on different neural structures, with the former involving linguistic brain areas in a left frontotemporal network and the latter depending on visuospatial brain areas in a bilateral parietofrontal network~\cite{HoudeMazoyer}. Experimental measurements of disjunctions $\alpha\vee\beta$ in human brains can be proposed along similar lines as suggested for compositions.

\begin{figure}[h!]
\begin{center}\includegraphics[width=0.5\textwidth]{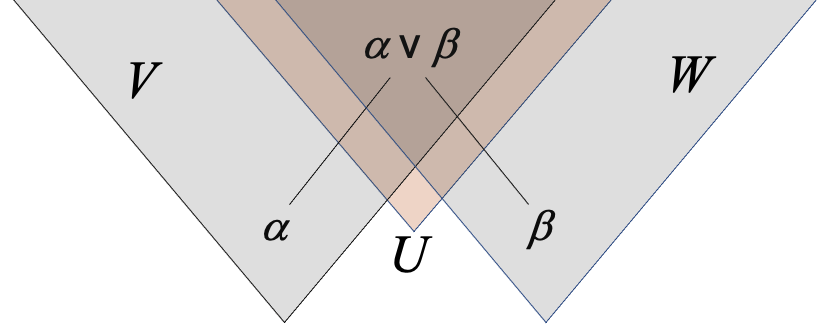}
\caption{Any concept $U$ containing $\alpha\vee\beta$ (the colored triangle) must contain the intersection of two concepts $V$ and $W$, each containing $\alpha$ and $\beta$, respectively.}\label{fig:disjunction}\end{center}
\end{figure}

Topologically, disjunction relates to concepts as follows: if a concept $U$ contains $\alpha\vee\beta$, then some instances of $\alpha$ and $\beta$ must trigger concepts $V$ and $W$ whose conjunction implies $U$ (Fig.~\ref{fig:disjunction}).

Following the analogy with measurement spaces, we assume disjunctions exist for all pairs of qualia, which leads to:

\begin{enumerate}

\item $\alpha\vee\beta$ is uniquely defined as the supremum of $\alpha$ and $\beta$---so the operation $\vee:\Q\times\Q\to\Q$ satisfies the laws
\[
(\alpha\vee\beta)\vee\gamma = \alpha\vee(\beta\vee\gamma)\ ,\quad \alpha\vee\beta=\beta\vee\alpha\ ,\quad \alpha\vee\alpha=\alpha.
\]
\item $\vee$ is \emph{continuous}~\cite{msnew}.
\end{enumerate}

In addition, the following properties are postulated, where $\impquale$ is the ``impossible quale'' (\cf\ section~\ref{sec:qualiaconcepts}):

\begin{itemize}
\item Minimality: $\alpha\vee\impquale=\alpha$ for all $\alpha\in\Q$.
\item Distributivity: $\alpha(\beta\vee\gamma)=(\alpha\beta)\vee(\alpha\gamma)$ and
$(\beta\vee\gamma)\alpha=(\beta\alpha)\vee(\gamma\alpha)$ for all $\alpha,\beta,\gamma\in\Q$.
\end{itemize}
The latter reflects the reading of $\beta\vee\gamma$ as ``$\beta$ or $\gamma$.'' 

The impossible quale enables the representation of compositions of qualia $\beta\alpha$ that are not supported by enough cognitive structure, such as if Alice experiences $\alpha$ and then Bob experiences $\beta$, in which case an indivisible experience of the composition may not exist, so we set $\beta\alpha=\impquale$. From the properties of measurement spaces it follows that $\alpha\impquale=\impquale\alpha=\impquale$ for all $\alpha\in\Q$. Hence, for mathematical convenience, we assume that every pair of qualia can be composed.

\paragraph{Insights.}

Both the AHA! moments of $\alpha\beta$ and $\alpha\vee\beta$ are basic forms of \emph{insight}. However, there are many other types of insight.

Consider a sequence of qualia $\alpha_n$ ($n$ a natural number) where $\alpha_n\le\alpha_{n+1}$ for all $n$, and assume $\alpha_{n+1}$ occurs ``later'' than $\alpha_n$ (in some underlying time). This sequence represents experiences becoming progressively less specific; equivalently, able to trigger more and more concepts: $N(\alpha_n)\subset N(\alpha_{n+1})$. If we view each $U\in N(\alpha_n)$ as a ``property'' of $\alpha_n$, the sequence resembles a computation where each step $n$ adds knowledge. This way of describing computations is typical in computer science, where instead of spaces of qualia there are spaces of states of computer programs: at each stage of a computation more properties become known about the thing being computed. For example, a computation of $\pi$ is an infinite sequence of increasingly precise approximations, \eg, ``3,'', ``3.1,'' ``3.14,'', ``3.141,'', etc. Each step corresponds to a stage of the computation, \ie, a state of a computer that runs a program for computing $\pi$.

While qualia are not ``states,'' we may consider the sequence $\alpha_n$ as an approximation to a ``limiting experience'' $\alpha$, the most specific quale which is less specific than all $\alpha_n$. This limit exists if the supremum $\alpha=\V_n \alpha_n$ exists in the specialization order of $\Q$. More generally, approximations need not be sequential. The appropriate concept is a \emph{directed subset} $D\subset\Q$, namely a nonempty set such that for all $\alpha,\beta\in D$ there is $\gamma\in D$ such that $\alpha\le\gamma$ and $\beta\le\gamma$: each pair of stages $\alpha$ and $\beta$ has a stage that improves upon both.

The limiting experience $\V D$ can represent various forms of insight, ranging from mathematical breakthroughs to social realizations or broad emotional states like optimism or pessimism. Perhaps the most abstract insight for a given living organism is the feeling that it exists as an independent entity; that is, the experience of a ``self.''

For such limiting experiences to exist, cognitive mechanisms must be available and able to integrate approximations into an induced quale. This process may run subconsciously until the final insight reaches awareness. Unlike direct sensory experiences, insights may result from hidden processing that synthesizes multiple stimuli, thoughts, or past experiences.

Similarly to composition and disjunction, measuring such new insights requires tailored experimental procedures that suit the cognitive apparatus behind each insight. I will refrain from discussing further experimental aspects, leaving such concerns to future collaborative research in experimental psychology.

Just as we imposed that $\Q$ supports disjunctions, let us assume it also has the supremum $\V D$ of any directed subset $D$, making $\Q$ a \emph{directed complete poset} (\emph{dcpo}). This completeness criterium implies that $\Q$ contains every quale that an ideally intelligent organism could compute---even if such an organism does not exist in practice.

Mathematically, due to the existence of disjunctions and the impossible quale, being a dcpo implies that $\Q$ is a \emph{complete lattice}, \ie, suprema exist for all subsets $S\subset\Q$. In particular, there is a unique largest element
\[
1=\V\Q.
\]
This quale is compatible with all nonempty concepts and may be interpreted as the subjective experience of ``just being,'' or ``the experience of the present moment,'' from which all other experiences arise as more specialized qualia $\alpha<1$.

However, $1$ does not necessarily represent the broadest experience of an individual being. If Bob's qualia form a subset $\B\subset\Q$, the supremum $\V\B$ might correspond to the least specific quale that Bob can experience---the quale ``Being Bob,'' in section~\ref{sec:qualiaconcepts} interpreted as his sense of self. This idea will be revisited in section~\ref{sec:agents} in relation to panpsychism and the combination problem.

\paragraph{Interdependence of qualia and concepts revisited.}

Measurement spaces are \emph{sober}. Translating to qualia and concepts, if a collection of concepts $F$ structurally resembles the open neighborhoods of a single quale---formally, $F$ is a \emph{completely prime filter}~\cite{stonespaces}---then there exists $\alpha\in\Q$ such that $F=N(\alpha)$.

This property has two key interpretations. First, it serves as a \emph{no-hallucination principle}: if the logical structure of concepts implies that a quale 
$\alpha$ should exist, then it truly exists. Second, it suggests that, given sufficient cognitive abilities, $\alpha$ corresponds to the experience of becoming aware of the entire concept set 
$F$. This highlights how stored concepts can influence subjective experience, reinforcing the interdependence of qualia and concepts.

Mathematically, sobriety implies that $\Q$ is a dcpo, meaning it supports the computational processes described earlier. However, since conscious organisms have finite cognitive ability, the subspace of qualia corresponding to their subjective experience may have more limited properties, for instance not being a dcpo. This topic partly motivates the discussion about ``conscious agents'' in the next section.

From this point forward, it will be assumed that $\Q$ is a sober space.

\section{Emergence of conscious agents}\label{sec:agents}

\paragraph{Agents in qualia space.}

Let 
$\B$ be the subset of 
$\Q$ containing all the qualia Bob can experience in his lifetime. As a topological space in its own right (with open sets 
$U\cap \B$ for all 
$U\subset\Q$), 
$\B$
inherits the 
$T_0$ property, so the interdependence of qualia and concepts applies to Bob.

If Bob has ideal cognitive abilities, he can experience the disjunction 
$\alpha\vee\beta$ of any two qualia $\alpha,\beta\in\B$, as well as the supremum 
$\V D$
of any ``computation'' described by a directed set 
$D\subset\B$. And we harmlessly assume that 
$\B$
contains the impossible quale 
$\impquale$. Then 
$\B$ is closed under arbitrary suprema in 
$\Q$, making it a sober space~\cite{msnew}*{Lemma 5.2}. This ensures Bob satisfies the no-hallucination principle mentioned in Section~\ref{sec:qualiaconcepts}. If Bob can also experience multiplications $\alpha\beta$, the mathematical structure of 
$\B$ closely resembles that of $\Q$.

Such subspaces, along with additional structure, provide the foundation for defining an \emph{observer}. Here we focus only on one aspect: how humans perceive the ``external world'' geometrically, for which we adopt a classicality principle:

\begin{principle*}[No quantum interference]
Human beings, and likely many other conscious organisms, are not directly aware of quantum interference effects, except in controlled settings like quantum experiments.
\end{principle*}

In terms of the lattice structure of 
$\B$, this translates to \emph{distributivity}~\cite{fop}:
\[
\alpha\wedge(\beta\vee\gamma)=(\alpha\wedge\beta)\vee(\alpha\wedge\gamma).
\]
From results in topology~\cites{bigdomainbook,stonespaces}, there is another sober space 
$S$ such that $\B$ can be identified with $\topology(S)$~\cites{fop,msnew}. Although $S$ is a \emph{fictitious} space---its points are not themselves qualia and cannot be directly experienced---it is mathematically significant. Unlike $\Q$, which lacks a conventional geometric interpretation (e.g., distances or curvature), 
$S$ belongs to a well-studied class of spaces in analysis and geometry: it is \emph{locally compact}.

\paragraph{Basic qualia.}

A classical computer processes information using finite bit patterns, which form a \emph{countable} set---one that can be enumerated using integers. From this, any computable infinite string, such as $\pi$, can be obtained through successive approximations. Similarly, physical measurements can be formally described (\eg, in a particle physics experiment), such that all measurement results ultimately stem from a countable set of basic measurements.

For qualia, however, the justification for a countable set of \emph{basic qualia} is different. As a working hypothesis, these should at least include qualia caused by direct sensory stimuli---such as visual or auditory experiences. Additionally, qualia related to basic insights or spontaneous thoughts may also be considered fundamental, including the AHA! moments of $\alpha\beta$ and $\alpha\vee\beta$.

The countability of basic qualia follows from the assumption that Bob's conscious experience unfolds as a sequence of discrete moments, each following from the previous one---without allowing Zeno-like sequences. For example, this assumption is supported by GWT. Thus, 
$\B$ should be a \emph{countably-based} lattice (or even finitely-based), meaning there exists a countable subset $E\subset\B$ such that, for every $\alpha\in\B$, we have $\alpha=\V A$ for some $A\subset E$. This means that the space $S$ is \emph{second-countable}, making it even more analogous to the familiar spaces of geometry and analysis.

\paragraph{Emergence of a geometrical world view.}

The emergence of $S$ highlights the geometric nature of perception. We experience a world filled with geometric shapes. A fully geometric reality seems to exist ``out there,'' and certainly theoretical physics has become heavily intertwined with modern geometry. Since this geometric structure cannot be directly linked to the topological structure of $\Q$, it is natural to ask how much of it stems from the space $S$. Do our brains process qualia in a way that constructs a geometric mental image?

\begin{figure}[h!]
\begin{center}\includegraphics[width=0.5\textwidth]{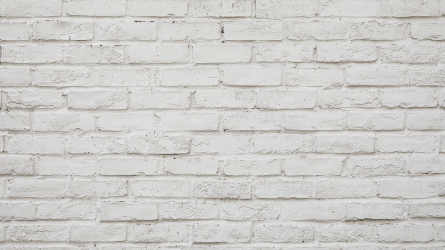}
\caption{When looking at a wall we receive information carried by photons reflected by the wall, which can only carry information about a region of the wall, no matter how small.}\label{fig:wall}\end{center}
\end{figure}

Consider a flat wall (Fig.~\ref{fig:wall}).
The photons reflected by the wall carry information about regions of the wall, not about its points (which physically do not exist, anyway, because no physical wall is a smooth surface). The stimuli produced in Bob's brain by such photons lead to his visual qualia associated to the wall. So, mathematically, it makes sense to consider a correspondence between Bob's qualia and open sets of the wall.

This example is accurate: an idealized smooth wall $W$ is a common geometric space (a flat Riemmanian manifold), and its set of open sets $\topology(W)$, itself equipped with an appropriate topology (the \emph{Scott topology}, commonly used in computer science), makes $W$ a sober space. This has the disjunctions required of a space of qualia (unions of open sets), and a multiplication given by intersection~\cite{fop}. Notably, the multiplication is commutative ($\alpha\beta=\beta\alpha$) and idempotent ($\alpha\alpha=\alpha$), reflecting the intuition that observing the same region of the wall twice does not yield new information and that the order of observations does not matter. These are familiar features of the ``classical world'' humans perceive.

\paragraph{What is it like to be Bob?}

Before the final section of this paper, devoted to another topological property of the space $S$, let us examine agents in qualia space through the lens of panpsychism. The space $\Q$, as a model of consciousness, is panpsychist because it does not restrict consciousness to specific organisms or physical systems. It does not even claim that consciousness arises ``everywhere,'' as $\Q$ does not contain physical space---geometric features like locations and distances can arise only within fictitious spaces like $S$ (\cf\ next section). Nevertheless, physical structures matter: as previously argued, the cognitive complexity of an organism endows its subspace of qualia with the mathematical properties that can explain the emergence of rich conscious experience.

Let us address the combination problem, returning to Nagel's question applied to variants of Bob such as ``What it is it like to be Bob seeing a flash of red light'' or ``Bob thinking that he thinks, therefore he exists.'' Each answer is a quale $\alpha\in\B$, possessing the indivisibility of an ``atomic'' conscious experience. In particular, Bob's greatest quale $\V\B$, proposed earlier to represent his abstract sense of self, has this property.

The point is that, despite the mathematical complexity of $\B$, the experience of ``being Bob'' in any given ``state'' remains a single quale, made possible due to a combination of more basic qualia via conscious or subsconscious computations with varying degrees of sophistication. Thus the emergence of observers in qualia space provides insight into the subject combination problem of panpsychism.

What if an organism, say Roach, has more limited cognition? Let $\C\subset\Q$ be its space of qualia. If Roach lacks sufficient computing power, $\V\C$ may not belong to $\C$, meaning Roach lacks a sense of self. Still, he can experience qualia like ``What it feels like to be hungry Roach'' or ``What it feels like to be threatened Roach.'' Thus, $\Q$ models the emergence of subjects with varying degrees of conscious experience.

\paragraph{Experiencing concepts.}

\begin{figure}[h!]
\begin{center}\includegraphics[width=0.5\textwidth]{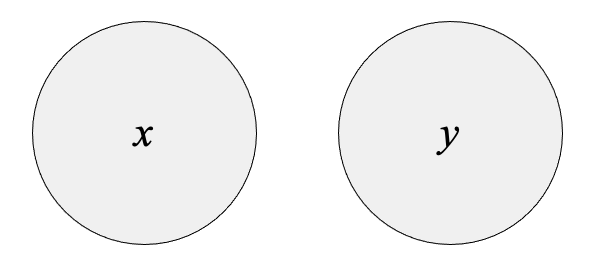}
\caption{In a Hausdorff space, distinct points $x$ and $y$ are separated by disjoint open sets (illustrated by the two disks).}\label{fig:haus}\end{center}
\end{figure}

\begin{figure}[h!]
\begin{center}\includegraphics[width=0.5\textwidth]{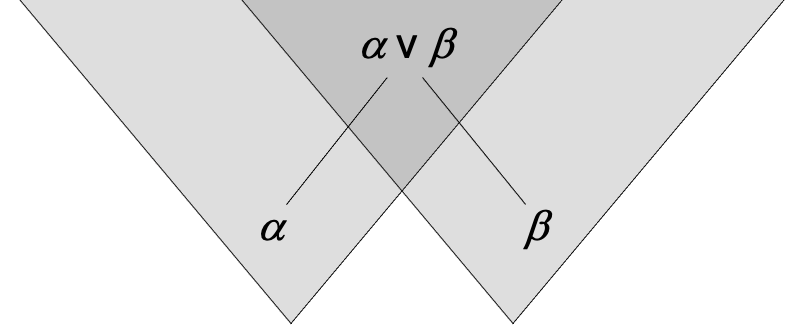}
\caption{Concepts containing qualia $\alpha$ and $\beta$ must intersect.}\label{fig:nonhaus}\end{center}
\end{figure}

In the wall example, the space $W$ corresponds to the fictitious space $S$, which is not only second-countable and locally compact but also possesses a well-defined geometry. While the discussion so far has not accounted for actual geometry, at least we can explain why $S$ should be a \emph{Hausdorff space}, meaning that any two distinct points $x,y\in S$ must be separable by disjoint open sets (Fig.~\ref{fig:haus}).
This property fails in $\B$ itself (as in $\Q$), because for all distinct qualia $\alpha,\beta\in\B$ and all concepts $U$ and $V$ containing $\alpha$ and $\beta$, respectively, we have $\alpha\vee\beta\in U\cap V$ (Fig.~\ref{fig:nonhaus}).

So why should $S$ be a Hausdorff space? Let us propose a mechanism based on the interdependence of qualia and concepts---specifically, how each concept $U$ has an associated experience. Let us formalize this by a function
\[
\Phi:\topology(\B)\to\B=\topology(S)
\]
where, for each concept $U\in\topology(\B)$, the quale $\Phi(U)$ represents Bob's subjective experience of recalling $U$. For example, if $U$ represents ``red,'' then $\Phi(U)$ corresponds to Bob's experience of evoking ``red.''

There is a cognitive basis for this. For example, extensive research on \emph{episodic memory} describes it as the ability to ``travel back in time''~\cites{Tulving2002,SRF22}, whereby recalling autobiographical episodes carries a subjective experience which is not confused with the experience of the actual episodes by a healthy subject. This type of experienced memory is concrete because it is inseparable from subjective time and the self of the experiencer, whereas the concepts which are represented by the open sets of qualia space are not necessarily related to either time or self. More generally, there is the notion of \emph{semantic memory}, concerning the storage and retrieval of knowledge in the brain, whose neural mechanisms have received wide attention~\cite{Kumar2021}.

However, current cognitive science is not aimed at explaining the mathematical properties of $\Phi$. Hence, notwithstanding that further experimental research should be guided by such an objective,
here a definition of $\Phi$ will be given based on intuitive and mathematical considerations. First, recall the asymmetry in the representation of concepts as open sets: intersections are finite, whereas disjunctions are arbitrary. Also, given the identification $\B=\topology(S)$, disjunctions of qualia in $\B$ are represented by unions in $\topology(S)$, and similarly for conjunctions. Hence, requiring the geometric logic~\cite{Vickers2014} of concepts to be respected by $\Phi$, we postulate:
\begin{eqnarray}
\Phi\bigl(\bigcup_{i\in I} U_i\bigr) &=& \bigcup_{i\in I}\Phi(U_i),\label{Phi1}\\
\Phi(U\cap V) &=& \Phi(U)\cap\Phi(V),\label{Phi2}\\
\Phi(\B) &=& S.\label{Phi3}
\end{eqnarray}
Since both $\B$ and $S$ are sober spaces, \emph{locale theory}~\cite{stonespaces} ensures the existence of a unique continuous function
\[
f:S\to\B
\]
such that $\Phi(U)=f^{-1}(U)$ for all $U\in\topology(\B)$. This represents the points of $S$ as qualia, suggesting that they can be experienced indirectly---perhaps $f(x)$ represents Bob's impression that $x\in S$ exists, such as when he ``believes'' a brick wall consists of points---so $S$ is not entirely fictitious after all.

Let us find an example of such a function $\Phi$. The rationale is that to have an experience of the concept ``red'' some non-red experiences must exist. After all, if we always wear red-lensed glasses, what basis will there be for distinguishing ``red''? (Notice that I am not referring to objective wavelength measurements.) This suggests defining
\[
\Phi(U) = \bigcup\left\{\alpha\in\topology(S)\mid\exists_{\beta\in\topology(S)}\ (\alpha\cap\beta=\emptyset\text{ and }\beta\in U)      \right\}.
\]
If $U$ is ``red,'' then $\Phi(U)$ is the disjunction of all the qualia sharply distinct from at least one ``red quale.''

The following is easy to prove:

\begin{proposition}
$\Phi$ satisfies the properties \eqref{Phi1}, \eqref{Phi2} and \eqref{Phi3}. Moreover, for each $x\in S$ the continuous map $f:S\to\topology(S)$ is given by any of the following expressions:
\begin{eqnarray}
f(x) &=& \left\{ y\in S\mid\exists_{\alpha,\beta\in\topology(S)}\ (x\in \alpha\text{ and } y\in \beta\text{ and } \alpha\cap \beta=\emptyset)\right\}\\
&=&\bigcup \left\{ \beta\in \topology(S)\mid \exists_{\alpha\in\topology(S)}\ (x\in\alpha\text{ and }\alpha\cap\beta=\emptyset)\right\}.\label{bigcupf}
\end{eqnarray}
And, defining $p:S\to\topology(S)$ by $p(x)=S-\overline{\{x\}}$, we have $f(x)\subset p(x)$ for all $x\in S$.
\end{proposition}

Let us examine the inclusion $f(x)\subset p(x)$. From the definition of $p$ (which in general is discontinuous) we obtain
\[
p(x) = \bigcup \left\{ \beta\in \topology(S)\mid x\notin\beta\right\},
\]
so it is immediate from \eqref{bigcupf} that if $\beta\in f(x)$ then $\beta\in p(x)$, proving the inclusion.

Now we have two maps
$
f,p:S\to\topology(S)
$.
Conceptually, $p(x)$ is the largest quale not containing $x$, but since $x$ is fictitious in the realm of subjective experience, the question of its membership in $\beta$ is ill-posed. Instead, Bob can only distinguish $\beta$ from some quale $\alpha$ containing $x$. This fundamental cognitive limitation implies $f=p$, which is equivalent to $S$ being Hausdorff:

\begin{theorem}
The equality $f=p$ holds if and only if $S$ is a Hausdorff space.
\end{theorem}

Thus, our analysis explains why Bob perceives the fictitious space 
$S$ as existing ``out there,'' a space that is second-countable, locally compact, and Hausdorff.

\section{Conclusion}

Qualia spaces are not a physicalist model because they introduce qualia without reducing them to known physics. Nor do they imply the opposite: qualia, as counterparts of quantum measurements, are potentially linked to nonlinear modifications of Schr\"odinger's equation. Regardless, qualia spaces are not dualistic. Developing them is valuable for the conceptual insights they offer, and the question of how they relate to physics can be pursued separately as a problem of \emph{representation} in other physical theories.

Conversely, measurement spaces, once endowed with mathematical structure encoding dynamics and quantum probability, are candidates for reconstructing quantum mechanics, and thus so are qualia spaces. Therefore, research into their interplay with quantum theory must continue.

Qualia spaces are capable of encoding complex structures because any theory of propositional intuitionist logic can be interpreted in the lattice of open sets of a topological space---see, \eg,~\ocite{IntLog2019}. In turn, topologies such as the concept topologies of Alice and Bob can be regarded as propositional logical theories, both intuitionist and geometric~\cite{Vickers2014}. It is natural to consider that two beings are ``similar'' if their logical theories of concepts are similar, with the highest degree of similarity being isomorphism of their concept lattices. At least Alice and Bob's concept logics should be similar enough to allow mutual understanding of significant subsets of their subjective experiences---a condition typically met by healthy humans.

Concretely, neural mechanisms including mirror neurons facilitate such understanding via empathic response. However, such machinery is ineffective when concept logics diverge too greatly. As Nagel's classic example illustrates, humans lack the empathic response needed to replicate a bat's experience, just as bats cannot comprehend human experience.

How does this impact the now-urgent question~\cite{Aida2025}---raised by the growing pervasiveness of sophisticated AI---of whether there can be artificial consciousness (AC)? From the perspective of qualia spaces, imagine a robot whose concept lattice $L$ is isomorphic to Bob's concept topology $\topology(\B)$. Then Bob and the robot might communicate and understand each other effectively. Due to sobriety, $\topology(\B)$ determines Bob's qualia space $\B$ up to homeomorphism---but so does $L$. The central question becomes: \emph{does the robot experience qualia modeled by a topological space $X$ homeomorphic to $\B$?} If yes, the robot may possess so-called strong AC. But baby Alice's first encounter with light suggests that color concepts can arise from physical events without involving qualia---\cf~section~\ref{sec:qualiaconcepts}. Likewise, the robot's concepts might have developed without subjective experience, making it a philosophical zombie---an instance of weak AC.

Thus, this framework suggests a \emph{necessary condition} for beings to have similar conscious experiences: their concept lattices must be similar. But this is not a \emph{sufficient condition}---two organisms with identical concept lattices are not necessarily both conscious just because one is. While the \emph{richness} of conscious experience may correlate with the complexity of a concept lattice $L$, reflecting an organism's cognitive abilities, the actual \emph{existence} of consciousness depends on the nature of the internal processes of the organism and their ability to \emph{create classical information} (as quantum measurements do), namely the information which is represented by its concept lattice.

For instance, Global Workspace Theory offers valuable insights into brain functions behind the richness of human consciousness. However, the point of the preceding arguments is that the level of consciousness of an artificial derivative of GWT---for instance a robot implementing the Conscious Turing Machine---should depend on its physical realization.

This raises a central question: which events in neurobiological systems generate classical information? At first sight, for quantum collapse to be relevant, it would seem that certain quantum superpositions must persist longer than typical decoherence times at biological temperatures. Consistently with this idea, Hameroff and Penrose~\ycite{HamPen14} propose that quantum coherence exists in neuronal microtubules, and that large-scale synchronization of collapse across neurons---\emph{orchestrated objective reduction (Orch OR)}---accounts for rich subjective experience. Notably, microtubules are found in all eukaryotic cells, including those of plants, animals, and fungi, suggesting the possibility of limited forms of consciousness even in simple organisms. Note that different models of quantum coherence in the brain may further elucidate coherence in microtubules, despite attributing consciousness to quantum computations---rather than collapse---and not addressing the measurement problem~\cite{AndrewBell2024}. In any case, it is fair to say that the role (or lack of it) of quantum measurements in neurobiology is yet to be clarified.

To conclude, intuitionist propositions can also be interpreted as qualia instead of concepts. For example, $\B$ is isomorphic to the topology of Bob's fictitious space $S$. Bob's quale $\V\B$---his sense of self---is the most abstract proposition within his intuitionist theory of qualia. The complexity of this theory mirrors the richness of the corresponding experience through a mathematical formulation that is far more abstract than Orch OR and not tied to neuronal structures. This interpretation also resonates with IIT's \emph{composition axiom}, which states that a system's consciousness relates to distinctions and overlaps among its subsystems.

\end{document}